\documentclass[useAMS, usenatbib, usegraphicx, referee]{biom}
\usepackage{amssymb}
\usepackage{mathtools}
\usepackage{graphicx}
\usepackage{booktabs, caption}
\usepackage{subcaption}
\usepackage{placeins}
\usepackage{mathrsfs}
\usepackage{bbm}
\usepackage[hyphens]{url}
\newtheorem{proposition}{Proposition}
\usepackage [english]{babel}
\usepackage [autostyle, english = american]{csquotes}
\MakeOuterQuote{"}
\DeclareGraphicsExtensions{.pdf,.png}

\title[De-Biasing the Bias]{De-Biasing the Bias: Methods for Improving Disparity Assessments with Noisy Group Measurements}

\author{Solvejg Wastvedt$^{1,*}$\email{wastv004@umn.edu}, Joshua Snoke$^{2}$, Denis Agniel$^{3}$, Julie Lai$^{3}$, Marc N. Elliott$^{3}$, Steven C. Martino$^{2}$ \\
$^{1}$Division of Biostatistics and Health Data Science, University of Minnesota, Minneapolis, Minnesota, U.S.A. \\
$^{2}$RAND Corporation, Pittsburgh, Pennsylvania, U.S.A.\\
$^{3}$RAND Corporation, Santa Monica, California, U.S.A.}

\begin{document}





\pagerange{\pageref{firstpage}--\pageref{lastpage}}




\label{firstpage}


\begin{abstract}
Health care decisions are increasingly informed by clinical decision support algorithms, but these algorithms may perpetuate or increase racial and ethnic disparities in access to and quality of health care. Further complicating the problem, clinical data often have missing or poor quality racial and ethnic information, which can lead to misleading assessments of algorithmic bias. We present novel statistical methods that allow for the use of probabilities of racial/ethnic group membership in assessments of algorithm performance and quantify the statistical bias that results from error in these imputed group probabilities. We propose a sensitivity analysis approach to estimating the statistical bias that allows practitioners to assess disparities in algorithm performance under a range of assumed levels of group probability error. We also prove theoretical bounds on the statistical bias for a set of commonly used fairness metrics and describe real-world scenarios where our theoretical results are likely to apply. We present a case study using imputed race and ethnicity from the Bayesian Improved Surname Geocoding (BISG) algorithm for estimation of disparities in a clinical decision support algorithm used to inform osteoporosis treatment. Our novel methods allow policy makers to understand the range of potential disparities under a given algorithm even when race and ethnicity information is missing and to make informed decisions regarding the implementation of machine learning for clinical decision support.
\end{abstract}

%



\maketitle


%

\section{Clinical Decision Support Algorithms in Health Care} \label{sec:introduction}
Statistical algorithms are frequently incorporated into health care decision tools used by health care providers and payers to predict risk for conditions or diseases, identify the need for early screening, make diagnoses or prognoses, plan treatment, and direct allocation of scarce resources such as transplantable organs \citep{celi2022}. 
While these algorithms, commonly referred to as clinical decision support algorithms (CDSAs), provide benefits, they may also contribute to healthcare disparities for different groups of individuals represented in the data. For example, an algorithm used to allocate disease management resources was found to contribute to racial and ethnic disparities in care because it incorporates prior healthcare use as a proxy for medical need \citep{obermeyer2019}. 

Disparities in algorithmic performance across groups are commonly referred to as \textit{algorithmic bias}, not to be confused with the traditional statistical notion of bias. To leverage the benefits of algorithms while minimizing the likelihood of incorporating inequities, methodological approaches have been recommended for assessing or correcting for algorithmic bias \citep{berk2021,hardt2016,johndrow2019}. These methods assume the availability of accurate information on the groups for whom we want to assess disparities, e.g., patient race and ethnicity \citep{cabreros2022}. However, in the health care setting, information on such characteristics is often missing, incomplete, or unreliable \citep{grantmakers2021}, whether due to privacy concerns, data collection limitations, or other factors. This lack of reliable information can produce biased (in the statistical sense) measures of algorithmic bias, which can lead to poor decisions regarding the impact of using CDSAs in health care settings. In particular, prior work has shown that disparity assessments relying on proxies for grouping variables may be statistically biased \citep{chen2019}. Organizations such as the United Kingdom Centre for Data Ethics and Innovation have published white papers calling for more attention to the statistical bias that arises from using proxy information to measure or mitigate disparities \citep{ukcdei2023}. In this paper, we provide methods to adjust for this type of statistical bias. In our case study, we focus on the common scenario of using imputed race and ethnicity information when evaluating CDSAs in healthcare.

\subsection{Algorithmic fairness in binary classification}
There has been increasing focus on the concept of algorithmic fairness in the statistics and computer science literature in the past five years. Broadly speaking, the field seeks to define measures of fairness, or equity, in the use of machine learning models. This includes, among others, considering inequities in the input data \citep{johndrow2019} or measuring disparities in the model outputs \citep{berk2021}. Fairness measures can be derived in the case of a binary output as well as continuous values or risk scores.

In this paper, we focus on disparate algorithmic performance based on a binary prediction. As we cover in more detail in Section \ref{sec:framework}, many fairness measures in this context rely on the confusion matrix, i.e., the 2x2 table comparing the actual binary outcomes ($Y$) with the predicted ($\hat Y)$ outcomes. Common fairness measures include differences in the group-level \textit{false negative rate} (FNR), \textit{selection rate}, and \textit{accuracy}. Table \ref{tab:nu} defines these and other metrics derived from the confusion matrix.

For the models we consider, binary predictions are determined by choosing a threshold and setting all predicted probabilities above or below the threshold to 1 and 0 respectively. While we focus primarily on measuring predictive performance disparities, other existing work seeks to mitigate these disparities by selecting optimal thresholds which minimize disparities on selected fairness measures \citep{hardt2016}. We discuss briefly how our work extends to mitigating disparities in the discussion in Section \ref{sec:discussion}. To define algorithmic fairness measures, we must select the group(s) across which we will measure algorithmic disparities. Groups are frequently defined by protected characteristics such as race and ethnicity, gender, age, or disability status. We focus in this paper on race/ethnicity as the grouping variable of interest, but our methods generalize to any groups or combinations of groups. 

\subsection{Use of proxies for protected group membership} \label{sec:introduction-bisg}
Self-reported data is the gold standard source of information on an individual's racial and ethnic identity \citep{oig2022}; however, often in healthcare settings only imperfect administrative information is available. As in the case of CMS administrative data, such information may have been collected using questions with a constrained set of response options that do not allow true self report \citep{filice2017}. These data could also be wholly or partly based on third party report. Even if self-reported data are available, these data can have high, often non-random, missingness  \citep{ng2017}.


Indirect estimation of race and ethnicity can be an efficient and valid way of filling in gaps in the data as self-reported information accumulates. 
For example, the Bayesian Improved Surname Geocoding (BISG) method is a commonly used methodology that can be used to produce accurate estimates of racial and ethnic group membership \citep{elliott2009}. The method combines information about a person's likely race/ethnicity from the U.S. Census Surname list with information about the racial and ethnic composition of the neighborhood they live in to produce a set of probabilities that a given person self-identifies with each of six mutually exclusive racial and ethnic groups: Hispanic, and non-Hispanic (NH) White, Black, Asian American or Pacific Islander, American Indian or Alaska Native, and multiracial. 

Using indirect, or proxy, information to infer race and ethnicity introduces additional statistical bias into estimates of algorithmic bias. Prior work on this topic is limited, but we follow some of the work proposed by \citet{chen2019} who derived an estimate of the statistical bias in one algorithmic fairness measure. We use the BISG and the modified BIFSG, an extension that utilizes first name information as well as surnames \citep{sorbero2022,branham2022} that is closely related to the original BIFSG algorithm \citep{voicu2018}, as our example methods for inferring race and ethnicity through proxy information, but the methods provided in this paper can be applied to the use of any proxy. 




\section{Statistical framework} \label{sec:framework}
Algorithmic fairness convention uses $A$ to denote group membership, which in our framework is a categorical variable with two or more levels. We let $Z$ indicate the covariates used to construct a proxy for $A$, and $P(A=a|Z)$ is the probability of membership in protected group $a$ given covariates $Z$. While we assume throughout this paper that $P(A=a|Z)$ is a probability, our methods can also be applied to a binary proxy variable such as that obtained by choosing a cutoff threshold for the probabilities or from a model that predicts binary values directly. However, probabilities are generally preferable to thresholded proxies because they retain more information.

Our goal is to predict a binary outcome $Y$ using a classifier $f$ that maps covariates $X$ to a prediction $f(X) \in [0, 1]$. We estimate binary predictions $\hat{Y}$ based on $f(X)$. In the common setting where a CDSA produces predicted probabilities, $\hat{Y} = I\{ f(X) > \tau\}$ is obtained by choosing a cut point $\tau$ for $f(X)$. We then assess fairness using common metrics for the performance of $\hat{Y}$ as a predictor of $Y$, calculating the metric for each category in $A$ and comparing across groups to identify predictive disparities. Example metrics include the false positive and false negative rates, positive and negative predictive values, and error rate. We can write a general form for these metrics using binary functions of $Y$ and $\hat{Y}$ and the group variable $A$. Definition \ref{def:nu} presents the general form used throughout this paper.

\begin{definition} \label{def:nu}
    Let $h_1(Y,\hat{Y}),h_2(Y,\hat{Y})$ be binary functions of $Y$ and $\hat{Y}$. Then $\nu(\hat{Y},a)$ is any performance metric of $\hat{Y}$ as a predictor of $Y$ that can be written in the following form:
    \begin{equation}
         \nu(\hat{Y},a) = \frac{E[I(A=a)h_1(Y,\hat{Y})h_2(Y,\hat{Y})]}{E[I(A=a)h_1(Y,\hat{Y})]}
    \end{equation}
\end{definition}

For example, the false negative rate can be written in this form with $h_1 = Y$ and $h_2=1-\hat{Y}$, giving $FNR(\hat{Y},a) = \frac{E[I(A=a)Y(1-\hat{Y})]}{E[I(A=a)Y]}$. Table \ref{tab:nu} shows other common performance metrics and corresponding choices of $h_1$ and $h_2$.

\begin{table}[]
    \centering
    \begin{tabular}{l|c|c|c}
        \multicolumn{2}{c}{Performance metric} & $h_1$ & $h_2$ \\
        \hline
        False negative rate & $\frac{E[I(A=a)Y(1-\hat{Y})]}{E[I(A=a)Y]}$ & $Y$ & $1-\hat{Y}$ \\
        False positive rate & $\frac{E[I(A=a)(1-Y)\hat{Y}]}{E[I(A=a)(1-Y)]}$ & $1-Y$ & $\hat{Y}$ \\
        Positive predictive value & $\frac{E[I(A=a)Y\hat{Y}]}{E[I(A=a)\hat{Y}]}$ & $\hat{Y}$ & $Y$ \\
        Negative predictive value & $\frac{E[I(A=a)(1-Y)(1-\hat{Y})]}{E[I(A=a)(1-\hat{Y})]}$ & $1-\hat{Y}$ & $1-Y$ \\
        Selection rate & $\frac{E[I(A=a)\hat{Y}]}{E[I(A=a)]}$ & $1$ & $\hat{Y}$ \\
        Error rate & $\frac{E[I(A=a)I(Y\neq\hat{Y})]}{E[I(A=a)]}$ & $1$ & $I(Y\neq\hat{Y})$ \\
        &&& \\
    \end{tabular}
    \caption{Choices of $h_1$ and $h_2$ allowing expression of common performance metrics in terms of Definition \ref{def:nu}.}
    \label{tab:nu}
\end{table}

Next, define a \emph{weighted estimator} $\hat{\nu}_W(\hat{Y}, a)$ of $\nu(\hat{Y}, a)$ as an estimator that substitutes estimated group membership probabilities $P(A=a|Z)$ for $I(A=a)$. As explained in Section \ref{sec:introduction-bisg}, group membership probabilities may be obtained from a model such as the BISG that uses covariates including name and geographic location to predict racial and ethnic self-identification. A weighted estimator incorporates the uncertainty in the group membership estimates into the group performance metric. For example, given observations $i = 1, ..., N$, the weighted version of the false negative rate is: $\widehat{FNR}_W(\hat{Y}, a) = \frac{\sum_{i=1}^N P(A_i=a|Z_i)Y_i(1-\hat{Y}_i)}{\sum_{i=1}^N P(A_i=a|Z_i)Y_i}$.

\section{Methods to address bias of weighted estimators} \label{sec:methods}

Weighted estimators are statistically biased when the group membership probabilities are not perfectly accurate, as described in \citet{chen2019}. However, the authors of this existing work derive a bias expression only for what they call the \emph{mean group outcome}, which measures average occurrence of the outcome $Y$ in a given $A=a$ group and thus does not directly apply to algorithmic fairness assessments. The mean group outcome can also be construed as the selection rate (rate of positive predictions) by replacing $Y$ with $\hat{Y}$, but the authors do not distinguish between $Y$ and $\hat{Y}$ in the paper and thus do not deal with more complex performance metrics encompassed by Definition \ref{def:nu} that include both $Y$ and $\hat{Y}$, such as the false positive and negative rates. The absence of any functions of $Y$ or $\hat{Y}$ in the denominator of the mean group outcome allows \citet{chen2019} to derive a simpler bias expression that cannot be used for other performance metrics. 

We substantially extend this work by deriving an expression for the statistical bias of any metric that can be written in the form $\nu(a)$. From our expression, we derive two sensitivity parameters that explain the sources of bias and can be used to estimate a plausible bias range for use by practitioners. We also propose a bound on the bias using quantities that may be estimated from observed data and delineate circumstances when this bound is valid.

\subsection{Bias of the weighted estimator $\hat{\nu}_W(a)$} \label{sec:methods-bias}

\citet{chen2019} showed that the bias of the weighted estimator of the mean group outcome is driven by dependence between $Y$ and $A$ conditional on $Z$. In Theorem \ref{thm:bias}, we derive a new formulation for the bias of any weighted estimator $\hat{\nu}_W(a)$ that is easily interpretable and connects more directly to the confusion matrix than the \citet{chen2019} construction. In simple terms, this theorem shows that the bias depends on the average group probability error for individuals in the cell(s) of the confusion matrix which contribute to the performance metric. More specifically, we show that the bias of any weighted estimator $\hat{\nu}_W(a)$ can be seen as a function of two quantities: error of group probabilities for observations where $h_1h_2=1$, and error of group probabilities for observations where $h_1=1$.

In this and following sections, we assume dependence of $\nu(a)$ and $\hat{\nu}_W(a)$ on $\hat{Y}$ and of $h_1, h_2$ on $Y, \hat{Y}$ and suppress the notation for conciseness. A proof of Theorem \ref{thm:bias} is given in the supporting information.

\begin{theorem} \label{thm:bias}
Let $h_1,h_2$ be binary functions of $Y$ and $\hat{Y}$. Let $\pi_a = P(A=a|Z)$ be the probability of membership in group $A=a$ given covariates $Z$. Define $\nu$ as the marginal equivalent of $\nu(a)$ such that $\nu = \frac{E[h_1h_2]}{E[h_1]}$, and let $\nu_W(a) = E[\pi_a h_1 h_2]/E[\pi_a h_1]$. Let $n_a$ denote sample size of each group such that $N = \sum n_a$. Then as $N \rightarrow \infty$, with $n_a/N$ converging in probability to a constant, the bias of $\hat{\nu}_W(a)$ converges almost surely to:

\begin{equation} \label{eq:bias}
    \hat{\nu}_W(a)-\nu(a) \xrightarrow{a.s.} \begin{multlined}[t] \frac{1}{E[I(A=a)|h_1=1]} \big\{ (1-\nu_W(a))\nu E[\pi_a-I(A=a) \big| h_1h_2=1] - \\ \nu_W(a)(1-\nu) E[\pi_a-I(A=a) \big| h_1(1-h_2)=1]\big\} \end{multlined}
\end{equation}.
\end{theorem}

For example, applying Theorem \ref{thm:bias} to the false negative rate, we have that the bias of $\widehat{FNR}_W(a)$ converges almost surely to the following:

\begin{equation} \label{eq:fnr_bias}
    \begin{multlined}[t] \frac{1}{E[I(A=a)|Y=1]} \big\{ FNR(1-FNR_W(a)) E[\pi_a-I(A=a)|Y(1-\hat{Y})=1] - \\ FNR_W(a)(1-FNR) E[\pi_a-I(A=a)|Y\hat{Y}=1]\big\} \end{multlined}
\end{equation}

Theorem \ref{thm:bias} highlights that the bias is driven by the relationship between error in the group probabilities and predicted and actual outcomes. Bias will be small when group probability error is low but also given small covariance between group probability error and outcomes. We demonstrate this relationship more directly in Web Appendix A with an alternative derivation that also connects our general form to the work of \citet{chen2019}.

\subsection{Sensitivity analysis approach to estimating the bias} \label{sec:methods-sensitivity}

The bias expression in Theorem \ref{thm:bias} cannot be estimated directly from observed data because it relies on the true group membership variable $A$. However, given a test set with $Y$, $\hat{Y}$, and group membership probabilities such as those provided by the BISG, the expression in Theorem \ref{thm:bias} enables a sensitivity analysis approach to characterizing a plausible range for the bias of any weighted performance metric. In Definition \ref{def:eps}, we denote two parameters from Equation \eqref{eq:bias} which we cannot estimate from test data as $\epsilon$ and $\epsilon'$. These parameters are easily interpretable as the average error of group probability estimation given particular combinations of the outcome $Y$ and prediction $\hat{Y}$.

\begin{definition}\label{def:eps}
    The sensitivity parameters $\epsilon$ and $\epsilon'$ are the expected values of group probability error given $h_1h_2=1$ or $h_1(1-h_2)=1$, respectively:

    \begin{align}
        \epsilon &= E[\pi_a-I(A=a)|h_1h_2=1] \\
        \epsilon' &= E[\pi_a-I(A=a)|h_1(1-h_2)=1]
    \end{align}
\end{definition}
For example, the false negative rate has $\epsilon = E[\pi_a-I(A=a)|Y=1,\hat{Y}=0]$ and $\epsilon' = E[\pi_a-I(A=a)|Y=1, \hat{Y}=1]$.

The marginal performance metric across groups, $\nu$ in Equation \eqref{eq:bias}, can be easily estimated using sample means, given a test set with $N$ observations, as 

\begin{equation*}
    \hat{\nu} = \frac{\sum_{i=1}^N h_1(Y_i,\hat{Y}_i) h_2(Y_i,\hat{Y}_i)}{\sum_{i=1}^N h_1(Y_i,\hat{Y}_i)}.
\end{equation*}

The group-specific weighted metric $\nu_W(a)$ is obtained similarly. This leaves one quantity in Equation \eqref{eq:bias}, $E[I(A=a)|h_1=1]$, for which we suggest two approaches. Most straightforwardly, this quantity could be treated as a third sensitivity parameter. However, for certain performance metrics, the choice of $h_1$ may allow reasonably precise estimation of this quantity from available sources outside the test data. For the false negative rate, where $h_1=Y$, the quantity is simply the expected probability of membership in group $a$ given a positive outcome, which for many clinical applications can be determined from prior studies of outcome rates for the groups in question, e.g., racial/ethnic groups, gender groups, etc.

To demonstrate the sensitivity analysis approach, we assume this third parameter can be obtained and focus on $\epsilon$ and $\epsilon'$. The bias is then estimated as follows:

\begin{equation}
    \widehat{Bias}[\hat{\nu}_W(a)] = \frac{(1-\hat{\nu}_W(a))\hat{\nu} \epsilon - \hat{\nu}_W(a)(1-\hat{\nu})\epsilon'}{E[I(A=a)|h_1=1]} \label{eq:bias_epsilon}
\end{equation}.

The sensitivity analysis is incorporated into estimation following the procedure outlined in \citet{zhao2019}, in which a nonparametric bootstrap is used to calculate the standard error of the weighted estimator $\hat{\nu}_W(a)$. For each replication, bias-corrected estimates $\hat{\nu}_W(a) + \widehat{Bias}$ are computed corresponding to a range of $\epsilon$ and $\epsilon'$ values selected by the practitioner. Means and $1-\alpha$ percentile confidence intervals are calculated separately for each $\epsilon, \epsilon'$ combination. Minimum and maximum mean values across all $\epsilon, \epsilon'$ combinations then comprise the \emph{bias-corrected plausible mean interval}, while the minimum low confidence bound and maximum high confidence bound comprise the \emph{bias-corrected sensitivity interval}. We use these terms to align with language elsewhere in the sensitivity analysis literature \citep{wolfson2010} and to differentiate from the standard confidence interval although, as \citet{zhao2019} show, the sensitivity interval has asymptotic coverage probability of at least $1-\alpha$.

We can use the fact that $E[I(a=a)]$ must lie between $0$ and $1$ to constrain the range of the sensitivity parameters $\epsilon$ and $\epsilon'$. Specifically, since the expected value of $\pi_a$ for the relevant combinations of $h_1$ and $h_2$ can be estimated, we have $E[\pi_a|h_1h_2=1]-1 \leq \epsilon \leq E[\pi_a|h_1h_2=1]$ and $E[\pi_a|h_1(1-h_2)=1]-1 \leq \epsilon' \leq E[\pi_a|h_1(1-h_2)=1]$. Removing $\epsilon$ and $\epsilon'$ that violate this constraint leaves a narrowed range of possible bias values. To further reduce computation time, note that the smallest (most negative) bias estimate for given $\epsilon, \epsilon'$ ranges corresponds to the smallest $\epsilon$ and largest $\epsilon'$. Likewise, the largest bias estimate results from the largest $\epsilon$ and smallest $\epsilon'$. For each bootstrap replication, practitioners need only calculate bias-corrected estimates corresponding to these maximum/minimum $\epsilon, \epsilon'$ values to obtain the bias-corrected plausible mean interval and sensitivity interval.

\begin{figure}
    \centering
    \includegraphics[width=.9\textwidth]{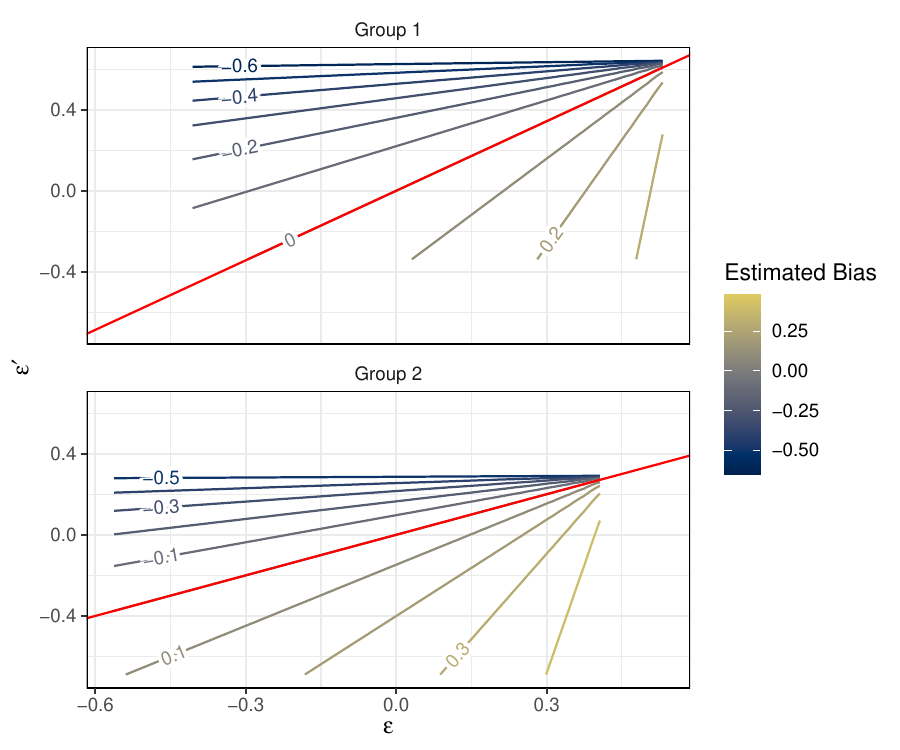}
    \caption{Example sensitivity analysis plot for the weighted false negative rate. The axes show $\epsilon$ (average error in group probabilities among observations for which $h_1h_2=1$) and $\epsilon'$ (the same among observations for which $h_1(1-h_2)=1$). Contour lines show estimated bias, with the diagonal line at the center of the figure indicating zero bias.}
    \label{fig:sens_demo}
\end{figure}

Sensitivity analysis results are visualized using a solid line for the plausible mean interval and error bars for the sensitivity interval, as demonstrated in the simulations in Section \ref{sec:simulations}. Practitioners may also explore the $\epsilon, \epsilon'$ parameter space using contour plots to find appropriate ranges, as shown using simulated data in Figure \ref{fig:sens_demo}. Bias is represented with colored contours using a diverging color palette to emphasize values farthest from zero. Note the application of the constraints on $\epsilon$ and $\epsilon'$: the parameter space is constrained separately for $A=0$ and $A=1$ according to the expected value of $\pi_a$ in the given $h_1$, $h_2$ groups.

\subsection{Bias bounds for selected metrics} \label{sec:methods-bound}

The bias in Theorem \ref{thm:bias} cannot be bounded using observed data without making additional assumptions. However, in certain scenarios we can exploit the fact that, as noted in the previous section, reasonable estimates for $E[I(A=a)|h_1=1]$ and $E[h_1]$ may be obtained from prior research. For example, when $h_1 \in \{Y,(1-Y)\}$, as in the false negative and false positive rates, the expectations we wish to estimate are the probability of membership in group $a$ given a certain outcome and the overall incidence of that outcome. In Proposition \ref{prop:bias_2}, we first present an alternative formulation for the bias that allows us to leverage estimates of these expectations. Since this bias expression still includes two terms not directly estimable from observed data, we then propose Assumption 1 that allows us to simplify the expression further and propose a bound on the absolute bias of the weighted performance metric.

\begin{proposition} \label{prop:bias_2}
Define $h_1, h_2$ as in Theorem \ref{thm:bias}. Let $\delta = E[(\pi_a-I(A=a)) h_1h_2]$ and $\delta^* = E[(I(A=a)-\pi_a) h_1]$. Then as $N \rightarrow \infty$, with $n_a/N$ converging in probability to a constant, the bias of the weighted estimator $\hat{\nu}_W(a)$ converges almost surely to:

\begin{equation} \label{eq:bias_2}
    \hat{\nu}_W(a) - \nu_a \xrightarrow{a.s.} \frac{\delta + \nu_W(a)\delta^*}{E[I(A=a)|h_1=1]E[h_1]} 
\end{equation}.
\end{proposition}

The terms $\delta$ and $\delta^*$ in Equation \eqref{eq:bias_2} capture the joint expectation of group probability error and membership in the groups for which $h_1=1, h_2=1$ and $h_1=1$, respectively. Thus $\delta$ represents a subset of the events included in $\delta^*$. Assumption 1 states that the expected absolute difference between estimated and true group probabilities when $h_1h_2=1$ is less than or equal to the absolute difference when $h_1=1$.

\begin{itemize}
    \item[] Assumption 1: $\big| E\{[\pi_a-I(A=a)]h_1h_2\} \big| \leq \big| E\{[I(A=a) - \pi_a]h_1\} \big|$, or equivalently, $|\delta| \leq |\delta^*|$.
\end{itemize}

For example, in the case of the false negative rate, where $h_1 = Y$, $h_2 = 1-\hat{Y}$, Assumption 1 compares estimation of group probabilities among false negative and positive observations. In Web Appendix B, we explore conditions that guarantee the validity of Assumption 1. One useful condition is the following: if the sensitivity parameters $\epsilon$ and $\epsilon'$, which capture conditional estimation error in the $h_1h_2=1$ and $h_1(1-h_2)=1$ groups, respectively, have the same sign, Assumption 1 is met (proof in Web Appendix B).

Under Assumption 1, we can use equation \eqref{eq:bias_2} and our estimates of $E[I(A=a)|h_1=1]$ and $E[h_1]$ to bound the bias. A proof of Theorem \ref{thm:bound} is given in the supporting information.

\begin{theorem} \label{thm:bound}
    Under Assumption 1, as $N \rightarrow \infty$ with $n_a/N$ converging in probability to a constant, the absolute bias of $\hat{\nu}_W(a)$ converges almost surely to a quantity obeying the following bound:

    \begin{equation} \label{eq:bound}
        | Bias[\hat{\nu}_W(a)] | \leq [1+\nu_W(a)] \big| 1- \frac{E[\pi_a h_1]}{E[I(A=a)h_1]} \big|
    \end{equation}.
\end{theorem}

In the supporting information, we give a full accounting of the conditions under which Assumption 1 is met and relate these conditions to the sensitivity parameters $\epsilon$ and $\epsilon'$. We note that the condition described above, in which expected group probability error has the same sign in both the $h_1=1,h_2=1$ and $h_1=1$ groups, is equivalent to requiring that $\epsilon$ and $\epsilon'$ have the same sign. 

While the bound holds for any $h_1, h_2$ given Assumption 1, as noted above we focus on metrics for which $h_1 \in \{Y,(1-Y)\}$ given our ability to estimate the denominator of Equation \eqref{eq:bound}. A similar argument applies to the error rate and selection rate, where $h_1=1$ and the denominator of Equation \eqref{eq:bound} can be estimated as the overall size of the group using population data. On a more technical note, use of the bound with the positive and negative predictive values presents additional problems because the bound is not sharp enough for practical application (Web Appendix C). This comes from the fact that for these metrics, the two terms in the numerator of the bias expression in equation \eqref{eq:bias_2} typically have opposite signs. Thus, taking the absolute value of each term as we do in the bound derivation causes the bound to widely overestimate the bias.

\section{Simulation study} \label{sec:simulations}

In this section, we explore the properties of the weighted estimator bias and our proposed bound. First, we compare uncertainty stemming from sampling error to uncertainty from bias to see which quantity dominates in scenarios practitioners may face. We then explore our sensitivity analysis approach, demonstrating how appropriate choices of $\epsilon$ and $\epsilon'$ change under varying data generation and accuracy of the group probabilities. Finally, we examine the sharpness of our proposed bound on the absolute bias. We find that modest sensitivity parameter ranges are sufficient to contain the true bias in even more extreme scenarios and that our bound provides a reasonable alternative to precisely quantifying the bias for certain metrics. The code to replicate our simulation studies is available at \url{https://github.com/swastvedt/weighted-algorithmic-equity}.

For the simulations in this section, we first generate $Z \sim N(-0.4, 1)$. We then generate the outcome $Y \sim Bin(p_y)$, where $p_y = expit(-0.2 + Z + X_1 + X_2 + X_3 + \beta_1 A)$, with $X \sim MVN((0,1,-1)^T, diag(0.5))$. The parameter $\beta_1$ represents the dependence between $Y$ and $A$ given the estimated group probabilities, which prior work has shown is a key factor controlling bias \citep{chen2019}. Taking the example of the BISG, $\beta_1$ represents the effect of race/ethnicity on the outcome not accounted for by the BISG inputs; in general, this effect can be thought of as extra predictive information in the group variable that is not captured by the inputs used to estimate the group probabilities. 

We simulate group membership by generating two sets of correlated probabilities, $P(A=a) = expit(Q_1)$ and $\pi_a = expit(Q_2)$ where the $Q$ are multivariate normal with mean $(Z, Z+\beta_2)$ and covariance matrix $\Sigma = 20*I_2 + \begin{pmatrix} 0&\beta_3 \\ \beta_3&0\end{pmatrix}$. The parameters $\beta_2$ and $\beta_3$ control the mean and AUC accuracy of the group probabilities, respectively. Setting $\beta_3 = 20$ represents near perfect covariance between true and estimated group membership, or an AUC of approximately $1$, whereas $\beta_3 = 0$ represents independence and an AUC of approximately $0.5$ (see Web Appendix C). This data generating mechanism assigns approximately 47\% of observations to group $A=1$. We split the data into training and test sets and create a model for $\hat{Y}$ from which we obtain predictions, dichotomizing with a threshold of $0.5$.

We connect our simulations to real-world BISG data to contextualize plausible values of $\beta_2$ and $\beta_3$. This real-world data consists of $7,007,871$ observations from the 2022 study of HealthCare.gov Health Insurance Marketplace enrollment data by \citet{sorbero2022}, from which we randomly sample $10,000$ observations. As described in Section \ref{sec:introduction-bisg}, the BISG produces probabilities of self-identification with each of six racial/ethnic groups. Our random sample is predominately NH White (55.1\%) with the following distribution over the other groups: Hispanic (20.9\%), NH Black (11.1\%), NH Asian American/Pacific Islander (10.1\%), NH American Indian/Alaska Native (0.7\%), and multiracial (2.1\%). We calculate $beta_2$ for each group as the difference between the true group proportion and mean BISG probability and $\beta_3$ as the group-specific area under the ROC curve (AUC) of the BISG probabilities. In our simulations, we mark the range of $\beta_2$ and $\beta_3$ values observed in this data. All analyses are done in R (version 4.3.0, R Core Team 2023). We focus on the false negative rate and group $A=1$; similar patterns hold for other metrics and $A=0$ except where noted. 

\subsection{Comparison of bias and sampling uncertainty} \label{sec:simulations-error}

In this section, we explore how sampling error compares to bias under varying sample sizes ($N_{sample} \in \{1,000, 2,000, \dots , 10,000$) with a population of size $N_{population}=50,000$. We also manipulate the conditional dependence parameter $\beta_1 \in [-0.5, 0.5]$. The ends of this range represent a fairly strong effect, since the magnitude is equal to roughly one standard deviation in the $X$ variables and half a standard deviation in $Z$.

\begin{figure}
    \centering
    \begin{subfigure}[b]{0.45\textwidth}
        \centering
        \includegraphics[width=.9\textwidth]{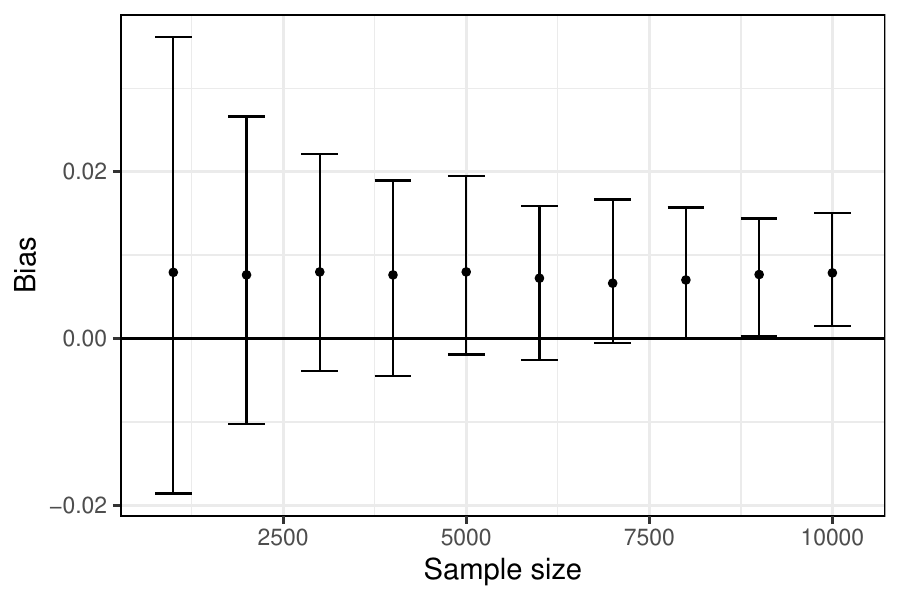}
        \caption{Comparison of sampling error and bias estimates at varying sample sizes with $\beta_1=0.25$.}
        \label{fig:sim_sampling_N}
    \end{subfigure}
    \hfill
    \begin{subfigure}[b]{.45\textwidth}
        \centering
        \includegraphics[width=\textwidth]{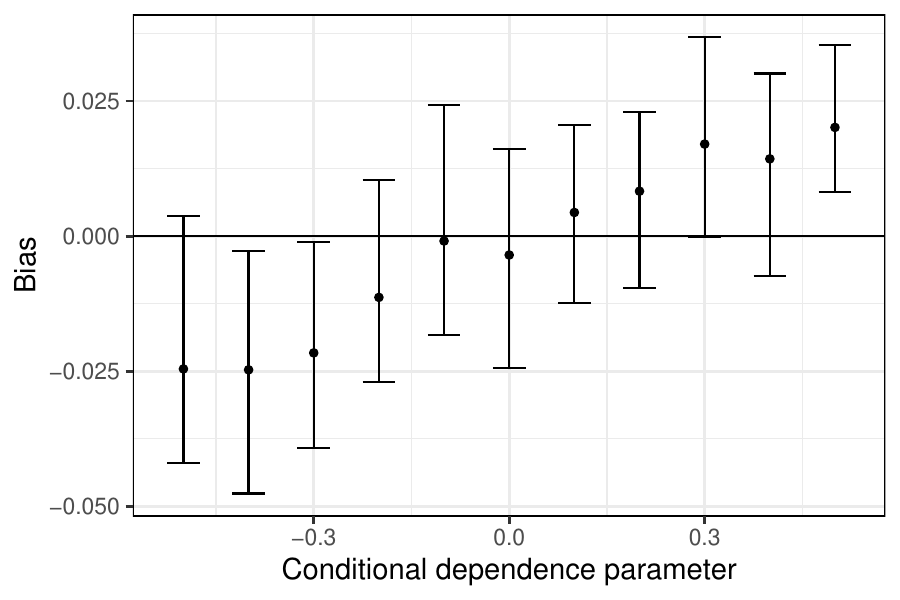}
        \caption{Comparison of sampling error and bias with varying levels of conditional dependence and $N_{sample}=2,000$.}
        \label{fig:sim_sampling_conddep}
    \end{subfigure}
    \caption{Comparison of bias and sampling error of the weighted FNR for group $A=1$ under varying sample size and conditional dependence. Dots and error bars represent means and 95\%-tile intervals, respectively, across $100$ replications of the simulation. Increasing sample size reduces sampling error with no impact on bias, while increasing magnitude of conditional dependence increases bias with no impact on sampling error.}
    \label{fig:sim_sampling}
\end{figure}

When we manipulate sample size while holding conditional dependence constant, sampling error (shown by the error bars) decreases with increasing $N_{sample}$, while the magnitude of the bias stays constant (Figure \ref{fig:sim_sampling_N}). Bias increases as magnitude of the conditional dependence parameter increases; this parameter has no effect on sampling error (Figure \ref{fig:sim_sampling_conddep}). Sampling error is more significant than bias under most scenarios we studied, with the exception of high levels of conditional dependence. This result indicates that in applications with limited sample size, such as that explored in our case study, uncertainty from estimating the bias may be a less pressing concern than that from sampling error. In practice, sample size also varies dramatically by racial and ethnic groups, as shown in the case study in Section \ref{sec:case-study}. NH White respondents make up roughly 55\% of the sample, while very small groups, such as NH American Indian and Alaskan Native often comprise less than 1\%. For small groups, the uncertainty due to sampling error will likely outweigh the bias.

\subsection{Choices of $\epsilon$, $\epsilon'$ for sensitivity analysis} \label{sec:simulations-sens}

In this section, we focus on estimating the bias using the sensitivity analysis approach of Section \ref{sec:methods-sensitivity}. This approach requires practitioners to choose ranges for the parameters $\epsilon$ and $\epsilon'$ as well as potentially the third sensitivity parameter $E[I(A=a)|h_1=1]$. Here we explore the impact of differing $\epsilon$ and $\epsilon'$ ranges on estimation of the bias, assuming, as in the previous section, that the third sensitivity parameter can be estimated from population data.

We examine sensitivity parameter ranges using three sets of scenarios. First, we manipulate dependence between $Y$ and $A$ conditional on $\pi_a$ as described in the previous section and quantified by $\beta_1$. Second, we shift the group probability means away from the true group proportions using the parameter $\beta_2$. Third, we manipulate the AUC of the group probabilities using the parameter $\beta_3$. For each set of simulations, we set the parameters not being manipulated to the following default values: $\beta_1=0.25, \beta_2=0, \beta_3=20$.

\begin{figure}
    \centering
    \begin{subfigure}[b]{0.7\textwidth}
         \centering
         \includegraphics[width=\textwidth]{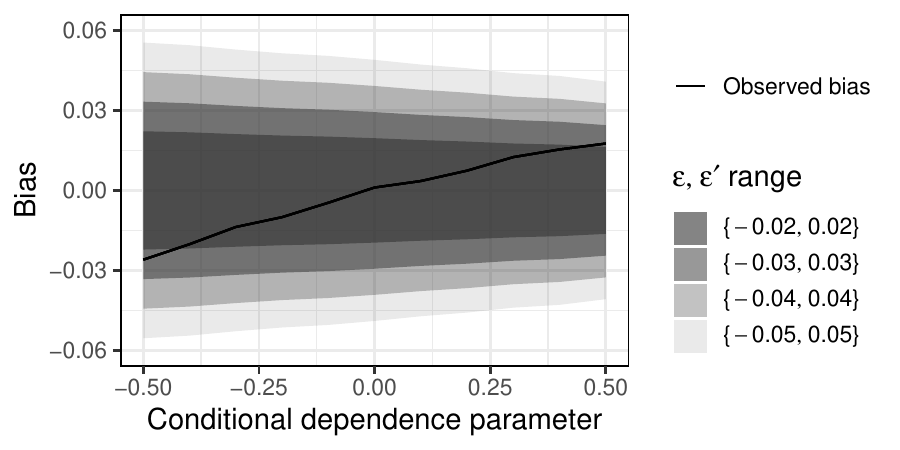}
         \caption{Varying $\beta_1$: Dependence between $Y,A$ conditional on $\pi_a$.}
         \label{fig:sim_epsilon_conddep}
     \end{subfigure}
     \hfill
     \begin{subfigure}[b]{.9\textwidth}
         \centering
         \includegraphics[width=\textwidth]{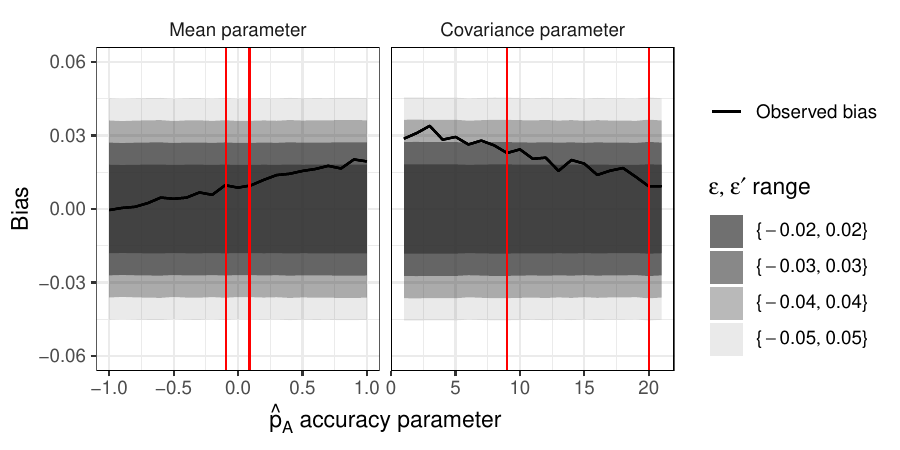}
         \caption{Left panel: Varying $\beta_2$ (difference in means of estimated vs. true group probabilities). Right panel: Varying $\beta_3$ (AUC of estimated group probabilities).}
         \label{fig:sim_epsilon_accuracy}
     \end{subfigure}
     \caption{Comparison of sensitivity parameter ranges under varying conditional dependence (top row) and accuracy of the group probabilities (bottom row). Shaded ranges show bias estimates for group $A=1$, and solid lines show observed bias, both averaged over $100$ replications. Vertical lines indicate ranges of the two group probability accuracy parameters ($\beta_2$ and $\beta_3$) observed in the real BISG data.}
     \label{fig:sim_epsilon}
\end{figure}

Figure \ref{fig:sim_epsilon_conddep} shows the ability of several different $\epsilon, \epsilon'$ ranges to contain the bias under varying conditional dependence. The smallest range contains the bias under moderate levels of conditional dependence, but a wider range (e.g., $\epsilon, \epsilon' \in [-0.03,0.03]$) is needed with more extreme conditional dependence. However, depending on true group proportions, this wider range may still represent a relatively small amount of group probability error. In our simulation, both groups are near $50\%$. For a group that represents $50\%$ of the population, a range of $[-0.03,0.03]$ accounts for scenarios where the estimated group proportion ranges from about $47\% - 53\%$, which is only 6\% relative error.

The accuracy of the group probabilities, as captured in $\beta_2$ and $\beta_3$, has a larger effect. Accuracy is directly tied to the values of $\epsilon$ and $\epsilon'$, which quantify mean group probability error in specific subsets of observations. Thus, as either $\beta_2$ or $\beta_3$ departs from perfect accuracy, represented by $\beta_2=0$ and  $\beta_3=20$, a larger $\epsilon, \epsilon'$ range is needed to contain the bias (Figure \ref{fig:sim_epsilon_accuracy}). To assess which simulations represent plausible scenarios for practitioners, we mark the range of $\beta_2$ and $\beta_3$ values observed in the real-world BISG data (vertical lines in Figure \ref{fig:sim_epsilon_accuracy}). These ranges represent accuracy across the six different racial/ethnic groups used in the BISG. While the BISG is strongly predictive of self-reported race and ethnicity for the four largest racial and ethnic groups in the U.S., accuracy is lower for the two remaining groups. \citet{elliott2009} find that AUC is 0.94 for NH Asian American/Pacific Islander, 0.93 for NH Black, 0.95 for Hispanic, 0.93 for NH White, 0.61 for NH American Indian/Alaska Native, and 0.77 for multiracial.

\subsection{Sharpness and validity of proposed bias bound} \label{sec:simulations-bound}

In Section \ref{sec:methods-bound}, we proposed a bound on the bias that can be used when Assumption 1 is met. Here, we investigate the sharpness of the bound for the false negative rate under varying levels of conditional dependence ($\beta_1$) and group probability accuracy ($\beta_2$ and $\beta_3$). Additional simulations in the supporting information show similar results for the false positive rate and demonstrate that while the bound does hold for other metrics (e.g., positive and negative predictive value), it is not the sharp bound desired for practical applications.

\begin{figure}
    \centering
    \begin{subfigure}[b]{0.7\textwidth}
         \centering
         \includegraphics[width=.9\textwidth]{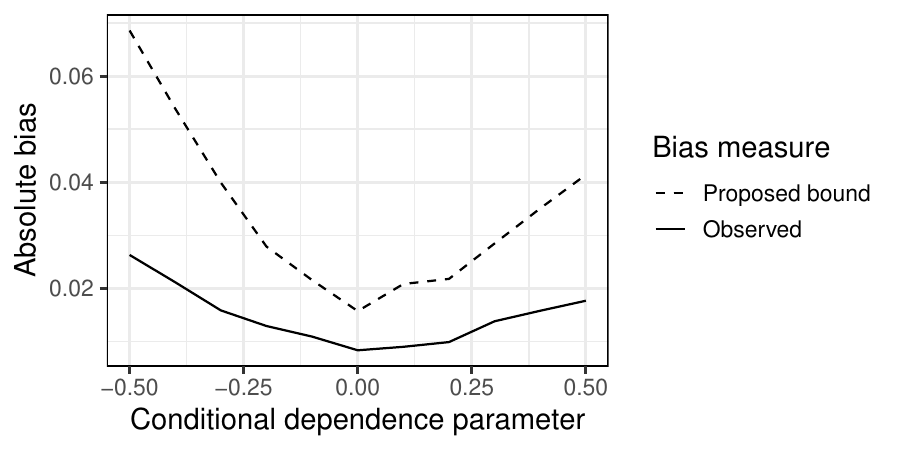}
         \caption{Varying $\beta_1$: Dependence between $Y,A$ conditional on $\pi_a$.}
         \label{fig:sim_bound_part1}
     \end{subfigure}
     \hfill
     \begin{subfigure}[b]{.9\textwidth}
         \centering
         \includegraphics[width=.9\textwidth]{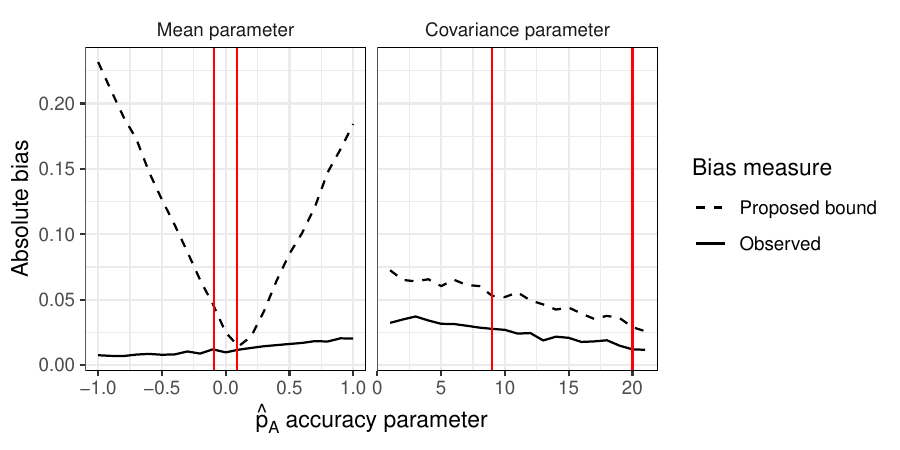}
         \caption{Left panel: Varying $\beta_2$ (difference in means of estimated vs. true group probabilities). Right panel: Varying $\beta_3$ (AUC of estimated group probabilities).}
         \label{fig:sim_bound_part2}
     \end{subfigure}
    \caption{Lines show the mean absolute observed bias and proposed bound on the absolute bias of the weighted FNR estimator for group $A=1$ across $100$ replications of the simulation. The proposed bound contains the absolute bias under all scenarios investigated.}
    \label{fig:sim_bound}
\end{figure}

As noted in previous sections, increasing magnitude of the conditional dependence parameter $\beta_1$ results in increased bias of the weighted FNR. Our proposed bound contains the absolute bias for all conditional dependence values investigated, although the bound is not particularly tight under high dependence (Figure \ref{fig:sim_bound_part1}). 

Manipulation of $\beta_2$ is of particular interest because our proposed bound relies heavily on the ratio of estimated to true $P(A=a)$ among observations where $h_1=1$. This ratio is driven by $\beta_2$; as $\beta_2$ diverges from $0$, the ratio diverges from $1$ (see Web Appendix C). Of the three $\beta$ parameters, $\beta_2$ also has the most significant impact on the sharpness of the bound: while the bound contains the actual absolute bias for all $\beta_2$ values simulated, the bound is not sharp enough to be useful for absolute $\beta_2$ greater than about $0.5$ (Figure \ref{fig:sim_bound_part2}, left panel). To understand this relationship, note that in formulating the bound, we substitute $|\delta| = |E\{[\pi_a - I(A=a)]h_1\}|$ for $|\delta^*| = |E\{[I(A=a) - \pi_a]h_1h_2\}|$. As the overall mean of the errors $\pi_a-I(A=a)$ increases with $\beta_2$, $|\delta|$ increases faster than $|\delta^*|$ because it draws on a larger subset of observations. Where $|\delta^*|$ includes only information from observations with $h_1,h_2=1$, $|\delta|$ draws from the broader group where $h_1=1$. However, the vertical lines show that the bound is still close to the true absolute bias under realistic values of $\beta_2$ observed in the BISG data.

Third, we manipulate the AUC between true and estimated group probabilities using $\beta_3$. Our bound contains the absolute bias for all values of $\beta_3$, with increasing AUC leading to a sharper bound (Figure \ref{fig:sim_bound_part2}, right panel). The bound is reasonable under the range of AUCs observed in the real-world BISG data (vertical lines).

\begin{figure}
    \centering
    \begin{subfigure}[b]{0.7\textwidth}
         \centering
         \includegraphics[width=\textwidth]{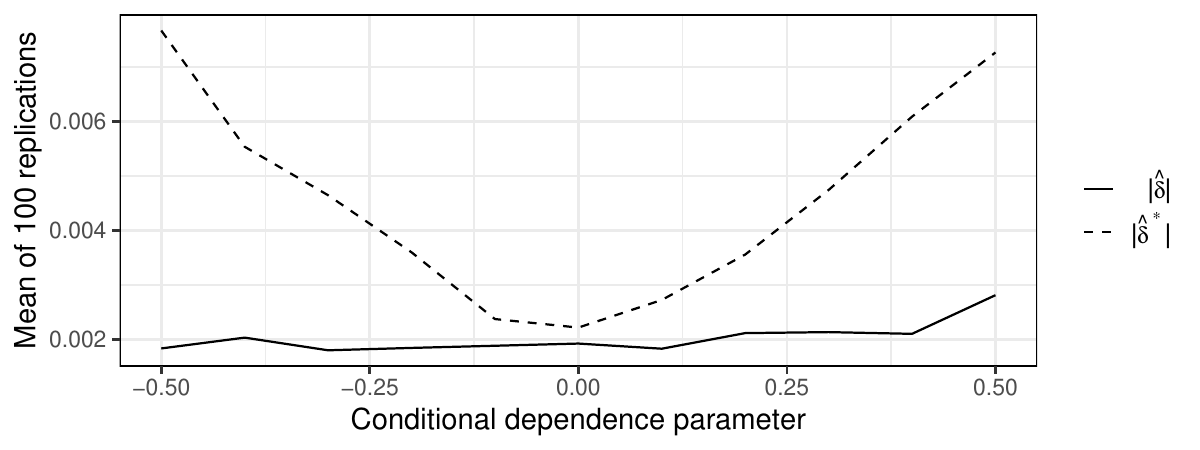}
         \caption{Evaluation of assumption under increasing dependence between $Y$ and $A$ conditional on $\pi_a$.}
         \label{fig:a1_part1}
     \end{subfigure}
     \hfill
     \begin{subfigure}[b]{.9\textwidth}
         \centering
         \includegraphics[width=\textwidth]{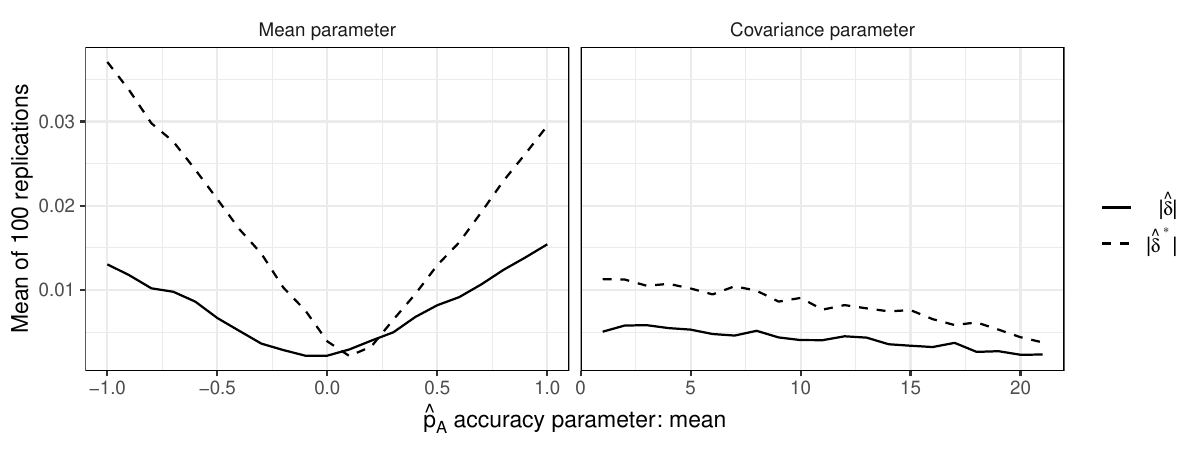}
         \caption{Evaluation of assumption under varying values of the group probability accuracy parameters: $\beta_2$ (left panel) and $\beta_3$ (right panel).}
         \label{fig:a1_part2}
     \end{subfigure}
     \caption{Comparison of left (solid line) and right (dashed line) sides of inequality in Assumption 1 for the FNR and group $A=1$ under each of the scenarios considered in Section \ref{sec:simulations-bound}. Lines show means of $100$ replications of the simulation. The assumption is met for all parameter values in the conditional dependence and covariance scenarios. In the mean shift ($\beta_2$) scenarios, the assumption is violated for a small subset of parameter values.}
     \label{fig:a1}
\end{figure}

Finally, we briefly examine conditions under which Assumption 1 is met, rendering the bound valid. Recall that Assumption 1 states the following: $\big| E[(\pi_a-I(A=a)) h_1h_2] \big| \leq \big| E[(I(A=a)-\pi_a) h_1] \big|$. We plot the left side of the inequality ($|\delta|$) vs. the right side ($|\delta^*|$) for each of the simulations above, estimating $\hat{\delta}$ and $\hat{\delta}^*$ using the sample data. Again focusing on the false negative rate and group $A=1$, we find that Assumption 1 is met for all $\beta_1$ and $\beta_3$ scenarios simulated, and for all except a small subset of $\beta_2$ scenarios (Figure \ref{fig:a1}).

\section{Case study: evaluation of the SCORE algorithm} \label{sec:case-study}
We replicate the SCORE algorithm \citep{cadarette1999} to show how one might use our bias estimation methods to improve understanding of racial and ethnic disparities in the performance of the algorithm. To replicate the algorithm, we selected a cohort comprising female Medicare fee-for-service enrollees with continuous enrollment from July 2016 to December 2017. Utilizing ICD10 diagnosis codes, we examined claims to identify instances of rheumatoid arthritis, fractures, estrogen use, obesity, BMI levels, and osteoporosis. Any enrollees who experienced a fracture following an osteoporosis diagnosis were excluded.

Our focus was on predicting the outcome—specifically, the occurrence of an osteoporosis diagnosis within six months of a fracture. We used a gradient boost model with predictors for diagnoses of rheumatoid arthritis, estrogen use, obesity, and BMI levels, age, and the total number of fracture claims. Given the predicted values, we choose a risk threshold for selecting a patient for further testing based on the expected utility approach proposed by \citet{pennello2016}. This approach balances the true positives and true negatives based on the overall prevalence in the population. Other approaches may use thresholds that meet the sensitivity requirements from early work on the SCORE algorithm \citep{cadarette1999}.

Using a sample test set of 100,000 observations, we estimate the FPR and FNR for each of the six racial/ethnic groups in the data using group probability estimates produced by the BISG and the modified BIFSG. We use five different values for the  $\epsilon$ parameters described in Section \ref{sec:methods-sensitivity}. The first two $\epsilon$ and $\epsilon'$ values are based on the differences observed between mean BISG and the modified BIFSG values and the true group proportions in the real-world data described in Section \ref{sec:simulations}. 

We assume these differences are the same for the subgroups of people who do or do not have osteoporosis, which is equivalent to assuming that the errors in the BISG model are not correlated with osteoporosis prevalence. We argue this is a valid assumption because we have no reason to believe osteoporosis is correlated with having an uncommon name for one's racial/ethnic group or being a minority in your Census block group. In addition to this, we estimate bias under $\epsilon$ and $\epsilon'$ parameters representing three progressively extreme levels of relative mean difference between estimated and true group proportions (5\%, 10\%, and 20\%). These differ from the BISG/BIFSG estimates because these assume the same relative mean difference for each group rather than varying by group.

For the FNR, $\epsilon$ and $\epsilon'$ quantify the difference between the mean group probability and the true population group proportion in the false negative and true positive categories, respectively. For the FPR, $\epsilon$ and $\epsilon'$ quantify the mean difference in the false positive and true negative categories, respectively. We use relative difference because the group proportions have significantly different magnitudes. We use a nonparametric bootstrap to account for sampling error, which is considerable for the smaller groups.

We also estimate the bias under two assumptions: uncorrelated and correlated errors, e.g., the errors in the group probabilities can be correlated with the errors in the predictive model, or they can be uncorrelated. For fairness measures such as FPR or FNR, correlated errors lead to more bias in the group-level estimates. For this model, we can likely assume that errors in estimating race and ethnicity based on names and locations are not correlated with errors in predicting osteoporosis. In other words, we do not expect that a correlation exists between an individual having an uncommon age, BMI, or medical history among those with osteoporosis and that individual having an uncommon name for their racial/ethnic group or being a minority in their Census block group.

We present results both assuming uncorrelated or correlated error, since in practice decision-makers could consider the potential disparities (or lack thereof) under either assumption. The results we display for correlated errors represent the worst-case scenario, where the errors are perfectly correlated (but the magnitude of the bias is fixed), so the reality might be somewhere between the two plots.

We estimate the third parameter in the bias equation using published rates of osteoporosis among women aged 65 and older \citep{noel2021} along with statistics on the total number of women aged 65 from different racial and ethnic groups \citep{u.s.censusbureauSEXAGE2021}. The third parameter is the proportion of each group within the population who have osteoporosis (for FNR) and who do not have osteoporosis (for FPR). The data originally had six racial and ethnic groups, but we drop two groups, NH Asian American/Pacific Islander and multiracial, because we could not obtain accurate estimates of the prevalence of osteoporosis for these groups.

\begin{figure}
    \centering
    \includegraphics[width=.9\textwidth]{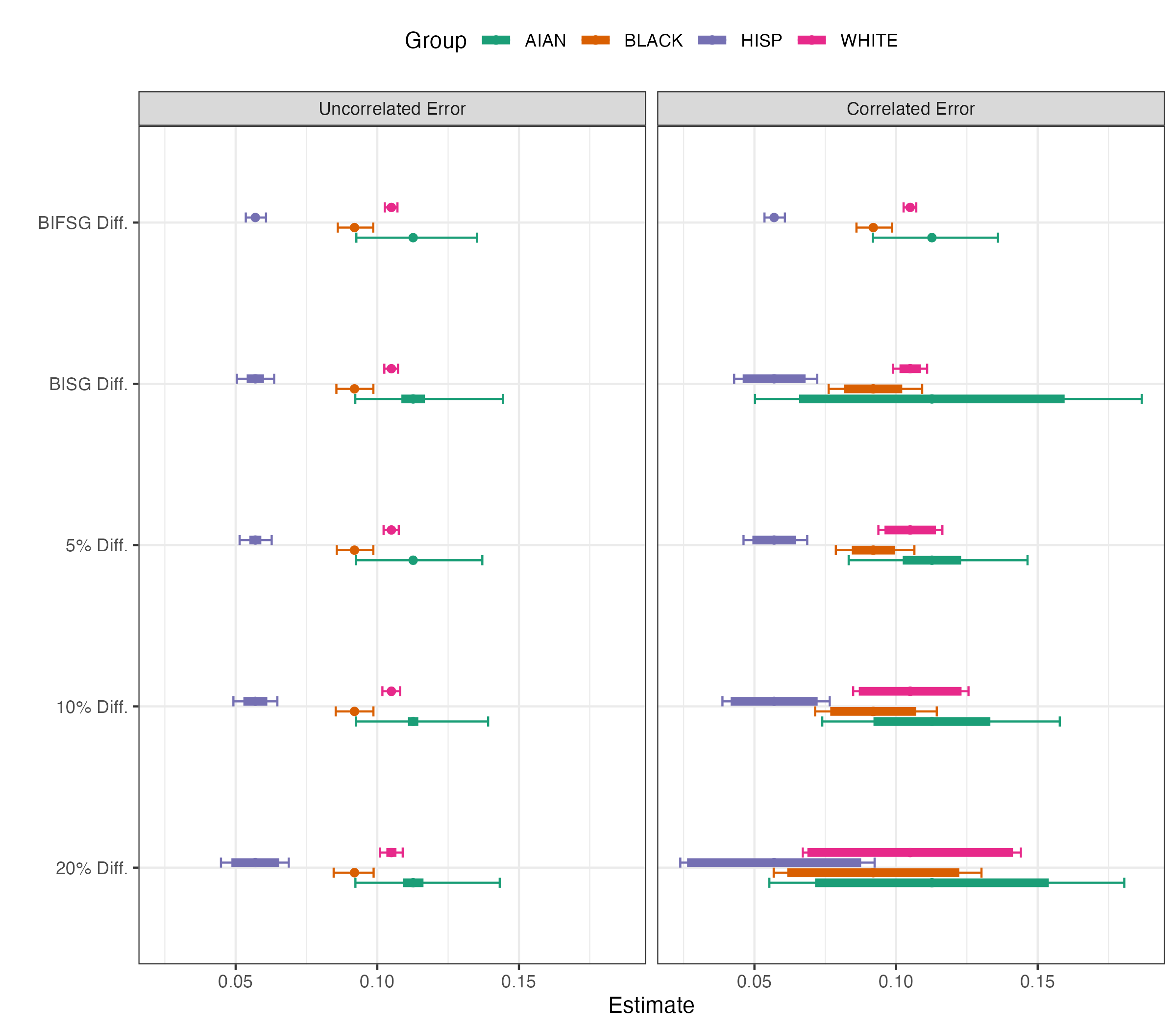}
    \caption{Estimated FPR intervals by racial and ethnic groups. Solid intervals show the estimated ranges for different levels of relative difference between the true group proportions and the mean BISG and modified BIFSG probabilities. Error bars indicate sampling error estimated using a nonparametric bootstrap. The two panels show estimates when assuming the group probability errors are either uncorrelated or correlated with the predictive model errors.}
    \label{fig:score_estimates}
\end{figure}

Figure \ref{fig:score_estimates} shows the FPR estimates at each level of assumed group probability error and for uncorrelated or correlated error. We see that if we assume uncorrelated error, there are some differences in group FPR rates, even when assuming high levels of error. For correlated error, the intervals for possible FPR estimates grow substantially if we assume the group probabilities contain significant error. For example, at 20\% relative difference in the mean probabilities from true group proportions, the bias intervals overlap for all four groups. In practice, this is likely an extremely high level of error in the group probabilities, depending on the proxy information used to generate probabilities. For error at the level of the modified BIFSG in the population, there is little bias in either case because the mean probabilities are very close to the true group proportions.

\begin{figure}
    \centering
    \includegraphics[width=.9\textwidth]{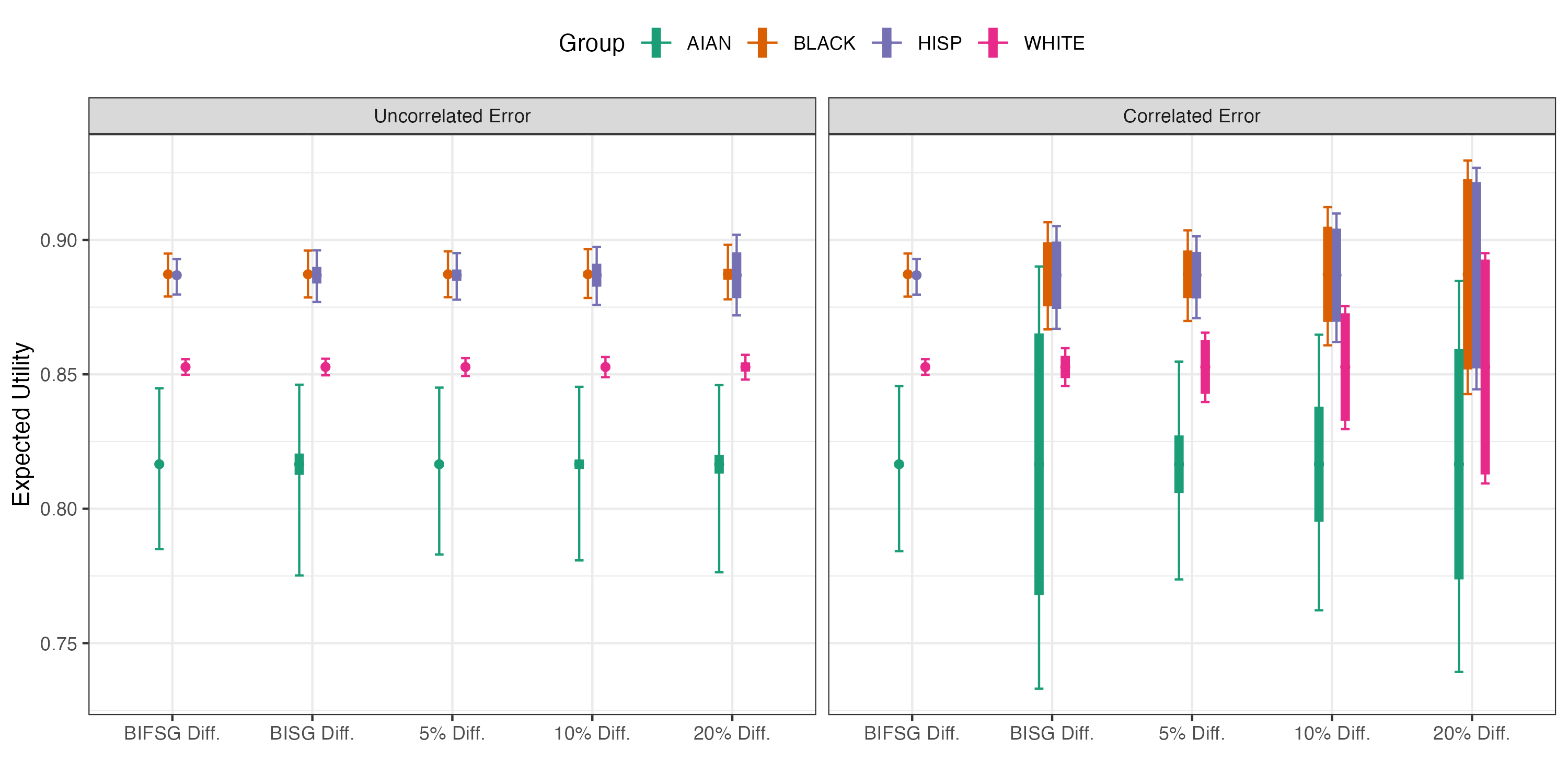}
    \caption{Estimated expected utility intervals by racial and ethnic groups. Solid intervals show the estimated ranges for different levels of relative difference between the true group proportions and the mean modified BIFSG and BISG probabilities. Error bars indicate sampling error estimated using a nonparametric bootstrap. The two panels show estimates when assuming the group probability errors are either uncorrelated or correlated with the predictive model errors.}
    \label{fig:score_eu}
\end{figure}

We combine FPR and FNR estimates into the measure of expected utility, which we used to select the overall risk decision threshold \citep{pennello2016}, i.e.,:
\[
EU = \underset{r \in (\frac{p_1}{p_0}, \frac{P_1}{1 - P_1})}{max} \hspace{2mm} p_0(1 - \tau_0)r + p_1(1 - \tau_1)
\]
where $p_0$, $p_1$ are the proportion of those without and with osteoporosis, $\tau_0$, $\tau_1$ are the FPR and FNR rates, $P_1$ is the average predicted value for positive cases.

Figure \ref{fig:score_eu} shows the results. Here we see that the NH Black and Hispanic groups have the highest expected utility from the algorithm predictions, and the NH White and American Indian/Alaska Native groups have lower utility. Interestingly, Hispanic individuals have high utility driven by low false positive rates, while NH Black individuals have high utility driven more so by low false negative rates. The disparities potentially shrink if we assume high levels of group probability error, correlated with the predictive model errors. Plots such as these showing bias-corrected measures for performance metrics and expected utility can help decision makers more accurately assess disparities in algorithm performance based on assumed levels of error in group probabilities.

\section{Discussion} \label{sec:discussion}

The novel bias expressions and estimation methods we propose in this paper allow practitioners to account for statistical bias resulting from using group probability variables in assessments of algorithmic equity. As such, our methods expand the possibilities for the assessment of algorithmic bias in CDSAs and in algorithmic fairness assessment in general. In common healthcare settings where information on protected groups such as race and ethnicity is not available, practitioners can use probabilities such as those provided by the BISG and our methods to gain insight into disparities in algorithm performance. 

The sensitivity analysis approach of Section \ref{sec:methods-sensitivity} permits estimation of the bias under a variety of assumed levels of group probability accuracy, while our bound on the bias proposed in Section \ref{sec:methods-bound} enables simpler assessment of group-level performance for certain metrics. Our bias estimation methods apply to any performance metric that can be written in the form of Definition \ref{def:nu}, thereby allowing practitioners to choose the appropriate metric for a given application. Finally, we demonstrate application of our methods to the SCORE algorithm to show how the bias expression and sensitivity analysis approach may be used in practice.

While our work encompasses many commonly-used performance metrics and substantially broadens the existing research in this area, we limit our discussion in this paper to binary classification scenarios in which predicted probabilities are dichotomized using a threshold. In many settings, the predicted probabilities themselves are of interest, and practitioners may prefer to use a performance metric such as the AUC. Further work in this area could extend to quantifying the bias of group probability versions of such metrics. 

The quality of assessments of disparities in algorithmic performance using our methods also depends heavily on the quality of the group probabilities used as inputs, although our work is the first we know of to quantify bias resulting from the use of probabilities. Many methods of obtaining group probabilities produce probabilities that are more accurate for some groups than others and may be substantially less accurate for the smallest groups. For example, the BISG has AUCs of 0.61 and 0.77 for the two smallest groups, compared to AUCs above 0.9 for all other groups. Combined with small group sizes, this lack of accuracy requires sensitivity parameter ranges that can represent a large relative difference in predicted vs. actual proportions and could limit the utility of the sensitivity analysis for these groups.

Our methods apply to group probabilities generated using any method; however, we use the example of the BISG throughout this paper because of its suitability for clinical applications. Practitioners using the BISG should pay special attention to the circumstances explored in our simulations that are relevant to this imputation method. As denoted in the figures in Section \ref{sec:simulations}, the generally high accuracy of the BISG means in most cases practitioners will only need to explore sensitivity parameter ranges on the narrower end of those we investigated. Regarding the bound, limiting scope to scenarios plausible under the BISG substantially improves sharpness since the more extreme $\beta_2$ scenarios we considered, in which the bound diverges substantially from the observed bias, fall outside the scope of BISG accuracy.

The results of our case study underscore the importance of utilizing accurate proxy information for generating group probabilities. The BISG, and in particular the modified BIFSG, are quite accurate methods for predicting group membership at the population level, so the bias is relatively low. As we saw, if the relative difference between the mean probabilities and true group proportions is higher, the bias in the disparity estimates may be quite substantial. It also underscores the importance of the assumptions concerning the correlation of model and group prediction errors. Practitioners should pay special attention to these assumptions when using our methods to determine the possible bias range, and we advise performing sensitivity analyses under multiple assumptions.

Finally, our methods extend to post-processing techniques that utilize group-specific thresholds to dichotomize predicted probabilities, such as equalized opportunity post-processing which balances the true positive rate \citep{hardt2016}. Post-processing can be used with the weighted estimator to find group-specific thresholds that balance the expected algorithmic performance given group probabilities rather than fixed values. Our methods could be applied after post-processing to estimate the bias of final assessments of performance disparities or incorporated during the optimization process to account for bias when finding group-specific thresholds. Future work should consider the impacts of each potential approach to accounting for bias during post-processing.


\backmatter



\section*{Acknowledgements}

The work described in this manuscript was supported by a grant from the Robert Wood Johnson Foundation (Grant ID: 79880) as part of its Pioneering Ideas program. It was also supported by funds from the RAND IS Incubator.

\vspace*{-8pt}

\bibliographystyle{biom} \bibliography{rwjf_paper}

\label{lastpage}

\end{document}